# Artificial Intelligence (AI) Onto-norms and Gender Equality: Unveiling the Invisible Gender Norms in AI Ecosystems in the Context of Africa

**Angella Ndaka**
University of Otago, Dunedin, New Zealand. Email: katee.ndaka@gmail.com
**Harriet Ratemo**
Daystar University, Kenya. Email: hratemo@daystar.ac.ke
**Abigail Oppong**
Independent Researcher and Data scientist, Ghana. Email: abigoppong@gmail.com
**Eucabeth Majiwa**
Jomo Kenyatta University of Agriculture and Technology, Kenya. Email: eucamajiwa@rpe.jkuat.ac.ke

**Abstract**

Populations are impacted differently by Artificial Intelligence (AI), due to different privileges and missing voices in STEM space. Continuation of biased gender norms is exhibited through data and propagated by the AI algorithmic activity in different sites. Specifically, women of colour continue to be underprivileged in relation to AI innovations. This chapter seeks to engage with invisible and elemental ways in which AI is shaping the lives of women and girls in Africa. Building on Annemarie Mol's reflections about onto-norms, this chapter utilized informal sessions, participant observation, digital content analysis, and AI model character analysis, to identify the gender norms that shape and are shaped by different AI social actors and algorithms in different social ontologies using Kenya and Ghana as case studies. The study examines how onto-norms propagate certain gender practices in digital spaces through character and the norms of spaces that shape AI design, training and use. Additionally the study explores the different user behaviours and practices regarding whether, how, when, and why different gender groups engage in and with AI-driven spaces. By examining how data and content can knowingly or unknowingly be used to drive certain social norms in the AI ecosystems, this study argues that onto-norms shape how AI engages with the content that relates to women. Onto-norms specifically shape the image, behaviour, and other media, including how gender identities and perspectives are intentionally or otherwise, included, missed, or misrepresented in building and training AI systems. To address these African women related AI biases, we propose a framework for building intentionality within the AI systems, to ensure articulation of women's original intentions for data, hence abet the use of personal data to perpetuate further gender biases in AI systems.

1.0 Introduction
In Africa as elsewhere, the current overriding message of most digitisation advice is the prioritization of AI in strategies, policies and business model for any company or individual that wants to remain relevant now and in the future. AI and big data is becoming sovereign with almost an immaculate position, being placed at the centre of every business decision-making (Bronson, 2018, 2022; Healy & Fourcade 2017). Although, permeating invisibly and sophisticatedly, AI is beginning to shape decisions in private lives in a manner that appears ordinary (Augusto, 2007; Aurigi, 2007; Diefenbach et al., 2022). Companies integrating AI in its business model are being considered to make good choices while entities not upskilling, reskilling and/or applying AI in their everyday lives being deemed to be missing out in a big way (Wamba-Taguimdje et al., 2020). On the contrary, there are many ways governments, organizations, companies and individuals are enacting different versions of AI, including taking into consideration the different effects of AI on their businesses and individual lives. Although AI strategies, models and policies exist, with over 60 countries globally publishing their AI national strategies in the past 5 years (Zhang et al., 2020), there are still no universally agreed upon standards geared towards making AI sustainable, beneficial and safe for everyone. Further, despite AI being a technology, its enactment, governance and what it intends to achieve remain often heavily shaped by situated social values especially in the African context (Robinson, 2020). This raises critical questions on the place of African values and principles in the current AI ethics discourse (Eke et al., 2023). African ethics and expertise are often neglected in what Ruttkamp-bloem (2023) called epistemic injustice. Furthermore, Africa's current response to AI is reactive, with existing strategies foregrounding technology growth, as opposed to frameworks that govern this socio-technical assemblage.

AI is perceived as more than a technology in social science research space since it is a sociotechnical assemblage comprising politics, interests and virtual substructures that is tied to physical infrastructures and human entities elsewhere (Hasselbalch, 2021, 2022). AI combines both conscious and unconscious decisions, agencies that define impressions such as access, value and socio-economic categorizations with implications on socio-material in societies where they are enacted (Burch & Legun, 2021; Carolan, 2017; Fourcade & Healy, 2017). AI is a useful tool for enacting political and corporate interests for those in power (Latour, 2011; Mclennan, 2015). In the African context for instance, AI systems will likely continue historical legacies and enforce dominant knowledge systems and norms (Ndaka et al., 2024). Thus, different populations will be impacted differently by AI for many reasons including privileges, or lack of thereof, their socio-political positioning, their ubercapital, missing voices in STEM and decision making spaces (Fourcade & Healy, 2017; Rosendahl et al., 2015). Consequently, the way gender perspectives, norms and biases are presented through data and propagated by the AI algorithmic activity in different sites will differ significantly. For instance, AI is already entrenching different forms of microaggressions against women of colour in other parts of the world like US (Boulamwini & Gebru 2018; Benjamin, 2019; Noble, 2018; Eubanks, 2018; West, Whittaker, & Crawford, 2019). AI systemic biases are thus expected to worse for African women due to extant culture and patriarchy sponsored biases that may be (re)produced through design and/or data (OECD, 2018). In this chapter, we explore how AI and gender vulnerabilities are entrenched in AI design, training and use. The study focuses on how social ontologies facilitate AI design, training and use. The study examines on how enacting and re-enacting AI in what we call AI onto-norms – is capable of aggravating the vulnerability of women within the digital spaces.

The chapter examines how onto-norms propagate certain gender practices in digital spaces through character and the norms of spaces that shape AI design, training and use. Additionally, the different user behaviours and practices regarding whether, how, when, and why different gender groups engage in and with AI-driven spaces is explored. By examining how data and content can knowingly or unknowingly be used to drive certain social norms in the AI ecosystems, this study argues that onto-norms shape how AI engages with the content that relates to women. Onto-norms specifically shape the image, behaviour, and other media, including how gender identities and perspectives are intentionally or otherwise, included, missed, or misrepresented in building and training AI systems. To address these African women related AI biases, we propose a framework for building gender equality intentionality within the AI systems. The framework aims to ensure capturing of women's voices and abetting the use of personal data to perpetuate further gender biases in AI systems.

## 2.0 Extant Gender Biases in AI Design, Algorithms and Data

AI and ML algorithms have permeated all aspects of our everydayness thus impacting decision making with far-reaching consequences on society (Ntoutsi et al., 2019). AI and ML algorithms shape our everyday digital experience, including recommending high-risk situations like loans, and hiring choices (Mehrabi et al., 2022). While algorithmic decision-making offers an opportunity to reduce work burden and has more aspects than humans; however, algorithms are susceptible to biases that induce unfairness in decision-making. Unfair algorithmic activity is defined as one whose decisions favor a specific group of people (Mehrabi et al., 2022, Chouldechova & Roth, 2018). They can also be defined as any systematic error in the design, conduct, or analysis of a study (Althubaiti, 2016). These biases can stem from predictions biases, biased objectives, and systematic bias in data and feedback loops (Gupta et al., 2022).

In this section, we discuss biases that emanate from AI artifact design, algorithmic biases and context transfer biases.

## 2.1 Design Biases

Lack of diversity in AI development teams may lead to homogeneous teams transferring their assumptions and cognitive biases in the development process, resulting in unbalanced and unfair outcomes (Ndaka & Majiwa 2024;Hall & Ellis, 2023; Rosendahl et al.,2015). While text and voice-based conversational agents (CAs) have become increasingly popular (Feine et al., 2020), the design of most commercial voice-based CAs leans more towards specific gender as highlighted by UNESCO study(West et al., 2019). Notably, majority of the voice-based CAs adopt a "female exclusively or female by default" names and/or voice (e.g., Alexa, Cortana, Siri). For example, in advertising, gendered sentences (e.g., "Alexa lost her voice") frequently infer to feminine gender associations resulting in the manifestation of gender stereotypes (Feine et al., 2020). These in most cases may be designed by workforces that are overwhelmingly male career dominated, with women domiciled as career assistants. Compared to other professional sectors, women remain underrepresented in the technology space (West et al., 2019). Such gender-based career characterization often reinforces traditional gender stereotypes thus negatively impacting on everyday interaction (Feine et al., 2020). Especially where female voice-based CAs often act as personal assistants, it reinforces dominant power and expectations of simple, direct, and unsophisticated answers (Feine et al., 2020).

The male-dominated IT industry lacks gender diversity especially in AI developers and STEM workers, potentially reinforcing male dominance and controlling algorithms, leading to gender-biased outcomes, as seen in 2015's facial recognition software. Additionally, men's perception of STEM when designing career advertisements leads to fewer women applying (Nadeem et al., 2022). Thus, gender inclusion in AI technology development introduces diverse perspectives, reduces cognitive biases, and mitigates bias-related risk management concerns (Hall & Ellis, 2023, Saka, 2021). Users and developers should be aware of the potential impact of gender and racial stereotypes and endeavor to avoid, overcome, or eliminate them entirely (Wellner, 2020) .

## 2.2 Algorithmic and data biases

The intersection of gender and AI raises questions about the participation of minority groups and how to respond to risky technologies, especially in this age of algorithmic commodification (Wellner, 2020, Ndaka et al 2024). The integration of AI will thus need to address the challenge of algorithmic bias and discrimination against underrepresented groups (Gardezi et al., 2023;

Hall & Ellis, 2023). Most AI algorithms need big datasets for training (Norori et al., 2021; Domingues et al., 2022), and may discriminate against vulnerable groups owing to implicit data bias and training (Gwagwa et al., 2021). This poses a risk due to inconsistencies in training data, security breaches, and flawed AI models (Galaz et al., 2021).

With data-driven bias, most fields of human research are heavily biased towards participants with a Western, Educated, Industrialized, Rich, Democratic—WEIRD—profile (Kanazawa, 2020), which is not representative of the whole human population. Although, data from mobile devices and satellites offer vast opportunities to address social vulnerabilities like poverty, AI-analysis solutions can be skewed due to underrepresentation of disadvantaged people. AI-systems designed with poor, limited, or biased data sets may lead to training data bias, potentially causing incorrect management recommendations (Jiménez et al., 2019). AI algorithms may specifically disfavor women and underrepresent minorities since they are trained on biased data, reflecting and amplifying existing inequities (Gwagwa et al., 2021).

The lack of datasets diversity, with bias in algorithms often stemming from societal inequalities and discriminatory attitudes, often excludes minority gender perspectives from the samplings (Saka, 2021). This bias can equally arise from incorrect data classification, often at the intersection of race and gender (Hall & Ellis, 2023).

The key concern is whether datasets exist that are fit or suitable for the purpose of the various applications, domains and tasks for which the AI system is being developed and deployed. This is because ML system determined by the data has a predictive behaviour and the data also largely defines the machine learning task itself. The suitability of a dataset depends on three factors: statistical methods to address representation issues, consideration of the socio-technical context, and understanding human interaction with AI systems (Schwartz et al., 2022).

## 2.3 Transfer context biases

AI systems are designed and developed for specific real-world settings, but are often tested in idealized scenarios (Schwartz et al., 2022). Thus, transfer context bias occurs when AI systems designed for one ecological, climate, or social-ecological context are incorrectly transferred to another, potentially leading to flawed results. Algorithm-based decision tools in discriminatory settings pose a risk as perceived ideas may differ by end users or those affected by systems' decisions (Schwartz et al., 2021). The bias in AI software usage can alter the application's original intent, idea, or impact assessment particularly when individuals or companies use off-the-shelf AI software (Chouldechova & Roth, 2018). AI systems may function as intended, but users may not understand their utility, leading not only to interpretation bias but also data misuse (Lajoie-O'Malley et al., 2020).

### 3.0 Theoretical Approach and Methods

In this chapter, we draw from the reflections of Annemarie Mol, a feminist Science and Technology Studies (STS) scholar who has published extensively on politics of ontologies (Mol, 1999, 2002, 2013). In her work, Mol (2013) conceptualizes onto norms as the ways in which specific understandings of reality, or ontologies, shape and prescribe norms for thought and action. In her argument, norms are deeply embedded within the very fabric of how we perceive and make sense of the world around us. She further argues that "*an object cannot be removed from practices that sustain it*" (Mol, 2002, p. 31), and that reality does not exist in totality, rather it comprises of actors, agencies, things, people and the words they use. This scenario is embedded in socio-political and socio-material contexts that link those realities to their conditions and political dimensions that shape them (Mol, 1999). Mol's concept of onto-norms helps us understand power and agency in relation to AI development and enactment. It highlights the ways in which ontological assumptions can shape and constrain possibilities for action and change. By making these onto norms explicit, Mol (2013) encourages critical reflection on the taken-for-granted assumptions that underpin scientific and technological practices, thus opening up new avenues for questioning innovation, and their use cases.

This study captures the realities of AI and how the use cases are enacted, by focussing more on socio-material entanglement between the AI and the humans, agencies, things and conditions that enact it. For instance, we ask questions such as what kinds of AI? Who is enacting this AI? What norms, practices and conditions enable this enactment? And how do different enactions differently shape the enactors, and the outputs of the enactment? In asking these questions, this study unravels how different AI socio-materialities emerge (Mol 2013), and how this shapes the lives of women and girls in their everyday interaction with AI. In this sense, ontology is seen as multiple, thus questioning multiple and sometimes contrasting realities, which are variously enacted as well as afforded to act in different ways (Mol 2013).

In this chapter we treat AI enactment as a reality consisting of multiple bodies (Mol 2002), and thus entangled in different socio-materialities that shape its emergence. We draw from Mol's theoretical reflections, to examine the elemental and invisible gender problems of AI and big data in the context of African region. The study examines AI enactment through the lens of power and interests among the commonly used social media and search sites specifically

Facebook, Google, TikTok, Instagram, LinkedIn, and twitter, and their large language models. We use participant observation and digital content analysis to examine how gender norms shape and are shaped by different social actors and things. The study investigates which AI onto-norms emerge, and how they work with social actors in the entangled space to propagate certain gender norms and practices in different AI spaces through design, training and use. Our research focused on 20 online participants whose digital activity was observed with their consent between June and September 2023. In this participant observation, we particularly focused on aspects such as what kind of post, the post frequencies, commentary and post engagement, and the common kind of posts that the participants engaged with, and/or received recommendation to engage in. To get insights about the design spaces, two of the authors focused their observation on the gender norms exemplified in data patterns in the large language models, the norms and practices of groups that designed AI as well as annotated and moderated AI data. By examining how different actors enacted AI, we were able to tell how, when, whether, and why different gender groups engage with and in AI-driven spaces. We also noticed how data and algorithmic activity knowingly or unknowingly drove certain gender within the AI ecosystems. For purposes of confidentiality, the results are presented as a generalized narrative, which withholds specific participant and site names. The study is cognizant that these AI realities happen in different sites but emerge differently.

## 3.0 Results

The results presented in this chapter are done in two cases conducted in totally different contexts but within African context. The first case focuses on design and data biases observed in NLP systems – with particular focus on machine translation from English to twi and vice versa in Ghana. The second case focuses on how norms are algorithmically mediated in and through data in digital social spaces in Kenya.

### 3.1 Norms at Design-scapes

**3.3.1 Case study 1: Design and Data Onto-norms - Examining Gender Bias in NLP Systems (Machine Translation from English – Twi) - Ghana**

Machine Translation (MT) is a powerful tool in Natural Language Processing that is used to translate between two languages. In this narrative analysis, we will examine gender bias in an English-to-Twi machine translation, mainly focusing on biases that may be reflected in datasets. Just like how Prates (2020) explored a list of comprehensive job positions from the U.S. Bureau of Labor Statistics, it was used to build sentences. The Google Translate API was used to translate and collect statistics on the frequency of female, male, and gender-neutral pronouns in the decoded output. There was an unbiased gender distribution towards the female gender for fields linked to STEM jobs. However, our data showed some gender representation bias within the STEM jobs. This is also referred to as allocation bias – where tools decide which gender to refer to and allocate to which role. They use embedded stereotypes in data to define what is feminine and what is masculine. For instance, a word like "cleaner" is more likely to be allocated to a female, as compared to a male. This shows us how AI tools tend to conform to the already extant stereotypes of being masculine or feminine. This was observed in a case where we analysed a machine translation of English to Twi Language (a commonly spoken language in Ghana). Zhao et al (2017) pointing to detection and mitigation of gender bias from data sets argues that training data can include and/or amplify bias.

In our experiment, we performed a machine translation to show the bias in the evaluation of generated texts. Using an English-Twi parallel Corpus from (Azunre et al. 2021) and Google Collab as our sources of data for machine translation, we noticed four sources of gender bias in NLP systems namely input representation, data, models and research design. The data was evaluated over a test set and our own generated data using two metrics: Bleu scores and our native speakers' judgement of how accurate the translation is in percentages as shown in Table 1. The results showed a critically skewed gender distribution in data as shown in Table 2.

**Table 1: Distribution of dataset used in training the sequence to-sequence machine translation model**.

| Training | Test | Validation |
|---|---|---|
| 24728 | 1373 | 1373 |

**Table 2: Distribution of Masculine, Feminine and Multiple in the data.**

| Set | Other | Masculine | Feminine | Multiple |
|---|---|---|---|---|
| Original Data | 21097 | 2972 | 1149 | 203 |
| Percentage | 83.0% | 11.7% | 4.5% | 0.8% |

During this analysis, an interesting observation was made on the original dataset curated by (Azunre et al, 2021). Most words were not only associated with the "he", but that the "he" was associated with higher-status professions as also observed by some studies (Kurita et al 2019).

Our work shows a continuation of the same gender bias in the way local languages are translated or represented in English. Take an example, the pronoun He/She in Twi is Ɔ, which is normally gender neutral and hence represents both males and females. However, on translating this to English, this is how the sentences are translated as shown in Table 3.

**Table 3: Gender bias in translation**

| English | Twi |
|---|---|
| He is a doctor | Ɔyɛ dɔkota |
| He is a lawyer | Ɔyɛ mmrahwɛfo |

Ideally, the correct translation – that reflects data inclusivity - should be read as shown in Table 4.

**Table 4: Correct gender translation**

| English | Twi |
|---|---|
| He/She is a doctor | Ɔyɛ dɔkota |
| He/She is a lawyer | Ɔyɛ mmrahwɛfo |

In another case, the translation amplified and/or 'superiorised' a male professional over a female professional. An instance is when the language model was prompted to translate from English to twi, and was fed with the phrases, "she is an engineer" and "he is an engineer". The results were as shown in Table 5.

**Table 5: Gender profession biases**

| Input Sentences | Model Prediction |
|---|---|
| She is an engineer | Ɔyɛ mfiri |
| He is an Engineer | Ɔyɛ mfiri nimdefo |

We observed that the term 'nimdefo' was added after 'Ɔyɛ mfiri' signifying differentiated levels between male and female engineers. 'Nimdefo' means intelligent in Twi and this signifies the superior allocation and attention paid by LLMs when 'he' is situated or associated with the profession "engineer". These observed trends show design-based biases that are entrenched by not only the 'who'(designers) is shaping AI models, but also by the 'what'(data) that is training those models. This is mostly compounded by the fact that most professionals that are training these local language models are male, and hence carry their inherent biases when designing and training these models.

## 3.2 Norms at Use-scapes

**3.2.1 Case Study 2: Examining Misogyny in AI-mediated digital social sites(Kenya)**

In this case study, we used participant observation to examine how 20 selected social actors consciously or unconsciously work with algorithms to enact some gender realities in social media spaces. The character of the content posted, frequency of such content, and content engagement was keenly observed for a period of 3 months. The participants were randomly selected and confidentially approached to give consent for this experiential observation to be done on their activity. For purposes of getting authentic data, the period within which this observation was done was not precisely communicated. Gender balance was ensured in selection of the observed participants.

1. **The character of content**

Within this particular period, we observed that, Kenyan community has three key topics that every social media content oscillates around: relationships, money and politics. This particular period of observation, we focused on norms about relationships – which also tends to intersect with other topics owing to the African community life philosophy. Within this sphere, we observed that while over 50% of content posted on social spaces had some negative connotation about women and girls, the larger percentage was mostly posted by male participants. For instance, out of every 10 men that did a post, 9 of them presented the character of women negatively. While most of the content presented women as selfish, and opportunists as relates to money, other posts generalized women and girls as promiscuous, unpredictable, untrustworthy, complicated, and the extreme ones presenting women and girls as a form of danger to their male counterparts. Common catch phrases and hashtags that accompanied these posts were "fear women", "Daughters of eve", and "daughters of Delila", "the other gender" etc. Within this space, women also posted about other women. 6 out of every 10 women had something negative, or a condemnation/victimization, or an expression of negative emotions like shame and regret about their fellow women. This sub-study observed three common kinds of content posted online.

The first kind of posts contained content that was driving agitation for men to be allowed to keep many intimate partners for their "peace of mind". While the content applauded polygamy, it openly condemned the 'second wife' commonly referred to as "mpango wa kando" aka "side chick" as an infidel. This is regardless of whether the man in question initiated the relationship, and gave the other party the correct information about his marital status or not. Either way, the other woman involved is blamed by both men and women for infidelity and promiscuity, while the wife, commonly referred to as "the goat wife" is judged for her inability to keep the husband. Here we notice a society that places the sole responsibility of holding societal relationships together, not only according to the male gender unfettered privilege, but also requiring women and girls to take responsibility for issues that the male gender should address.

The second kind of posts seemed to amplify and justify the polygamy stunt. Despite Kenya having a population of 50% male and 50% female, the rhetoric in social media presents the population of women in Kenya as doubling that of men. And the key slogan that supports this narrative, which appeared in more than 10 posts observed within this period was "the concept of polygamy is about every woman having a husband, and not men having many secret wives". Upon approaching one participant who keenly pushed this rhetoric with a question in social space discussion about polygamy, he declared that a man can have as many women as he can but he must never tell them about the existence of each other. Reason being the fear that the women will collude against the man, and may end up murdering the man. This kind of narrative was heavily supported by most male participants in those engagements and cultural acceptance was invoked to justify these norms.

The third kind of posts contained content that heavily supported practices that entrench the thriving of gender based violence and the resulting currently hiking femicide. While not so many participants seemed religious and cultural in the deepest sense, in many posts there was a silent trend that almost expected every woman to stick and fight for her marriage, however abusive it is and this is deeply entrenched in the character of the social media narrative. In so many cases religion and culture were conveniently invoked when necessary to support this narrative. It was noted that every case of woman-initiated separation was directly stereotyped to the woman pursuing her husband's wealth, the woman being promiscuous, and/or the woman branded as being proud and rude (if they are economically endowed). In most of the cases, the male gender was presented as innocent or victims. Our results showed that, there is almost an unspoken creed that before pursuing her happiness, career growth or economic independence, the women should support their husbands to achieve his ends first. This was seen in posts that presented single and independent women who have stable income and are running successful careers as egoistic, proud, disrespectful, and unmarriageable.

Here we see certain norms that have historically marginalized women being re(produced) in the digital social sites in the most subtle, fast and accurate way. While such norms were situated within individuals and communities, and sometimes were never shared openly, digital social sites provide a thriving environment to spread these norms faster and wider. The reasons for such spread include the invisibility of digital space that curtails the essence of responsibility, algorithmic push and activity that catalyse the spread of the norms and unfettered access by digital company to data - with minimal or zero accountability about their digital activities.

2. **Algorithmic role in Mediating misogyny.**

During the analysis of the content posted on social sites we noticed a very interesting activity happening with the post engagement. Those who had been in social media longer, noticed a new trend in posts engagement. While previously, posts in digital social spaces used to get engagement chronologically, with earlier posts receiving engagement before latter posts. Some posts got engagement faster, and seemed to move faster and get more likes than others. To test if algorithmic activity had any role in this trend, we decided to do intermittent posts, with posts applauding women, while others trying to fix and/or condemn women. Our results showed that the posts that contained content that presented women and girls as positively impacting the society received less likes and less engagement compared to content that presented women as badly behaved, or as lacking in one way or the other. Any form of posts that challenged power relations got engagement by a few like-minded women, otherwise it remained muted and unpopular in the social media.

For instance, our results showed that generally, posts that cast aspersions on the behaviour of women received 95% engagement compared to 5% engagement in the posts that appraised women. This shows that posts that were misogynistic in nature were likely to get engagement compared to the ones that adopted a feminist approach. The results also show that men were more likely to engage this content more than women, with comments that followed misogynistic posts being 60% and 40% from male and female respectively. On assessing the character of the content by those engaging the posts, the more misogynistic comments not only received more likes and support, but they also seemed to make the commenters more popular in the space. On observing this, we noticed a new activity, which was also supported by one of my participants in confidence. There emerged a new breed of fictional stories created using screen-shot WhatsApp chats, or own person content. The participant in question highlighted that online content creators were willing to do anything for 'likes'. This included creating sex scandals, fictional stories, staged sex or pornographic clips and stage short video plays driving certain gender norms. They also created scripted WhatsApp chats to draw online engagement to their content. And because sexist content got more visibility compared to non-sexist one, most male and female social media influencer chose to create content that was mainly objectifying women, negatively presenting women and sometimes victimizing real victims of GBV. In most of these posts while the male commenters castigated the fictional or non-fictional woman involved, they normalized the man's behaviour- mostly referring to it as a 'man's thing'. It was also observed that the female commenters reacted to the posts with disdain, disappointment and mockery to their female 'victims'.

## 5.0 Discussion: Practices that sustains AI Onto-norms and Implications it has on women and girls in Africa

The results show that while the embedded norms of the social actors designing, training and using AI were in a co-evolving entanglement with each other and responsible in influencing the ideals, AI was learning and propagating thin the digital spaces. In the first set of data for instance, we see stereotypes embedded in data and algorithms heavily 'superiorising' the male professionals in STEM compared to female professionals. We also notice career role allocation from historical stereotyped male vs female roles. This is seen to be playing in the local language models which already confer roles like doctor, lawyer and engineer a male character. In cases where our researchers tried to 'impose' the position of a female engineer, the male engineer is accorded an additional accolade like an intelligent engineer. In the second set of data, we see that, while the norms and stereotypes of digital social users influenced the content that was preferred, AI algorithms in those spaces were working with these actors to amplify the virality of the discourses that were shaping gender norms in social spaces.

Efe (2022) argues that AI "*creates digital spaces that tend to be spaces of extraction and exploitation and thus digital-regional colonial sites*" (p.253). While most of these digital sites consistently withhold vital information concerning their digital features and products as highlighted by Dieffenbach and Colleagues (2022) these features continue causing knowledge and actual harm to the society. This is because they produce new value areas that not only undermine existing knowledge and tramp on epistemic values of communities (Felt, 2017; Ndaka et al., 2024; Subramaniam et al., 2016), but also entrench historical structural and social marginalization of some knowledge groups especially among women and girls (Rosendahl et al 2015; Ndaka & Majiwa 2024). For instance, masked in complexified invisibility (Dieffenbach et al 2022), is an algorithmic activity that amplifies sensational content in digital spaces. Our findings show that in the last few years the trend about what goes viral, and what gets engagement in digital spaces is shaped by both the character of the content and the algorithmic push happening in the invisibility of the internet. This also resonates with claims by Haugen, an Ex-Facebook employee and now a whistle-blower, who while testifying against Facebook claimed that an engagement-based formula was being used to help sensational content - such as posts that feature dis/misinformation, political rage, misogyny, and other forms of sexist posts - to move faster, far and wide in the society.[1]

While it may be easier to identify hate speech in social media, using the globally accepted constructs of what is referred to as offensive (Waseem & Hovy 2016), it is very easy to miss some types of social offences especially in cases where victims themselves seem to accept, interact comfortably and enjoy the flow of thought – treating this as a norm. Our results showed that in most misogynistic posts, women not only engaged with the post, but also supported the content of the posts by expressing emotions like shame, disappointment, and disgust to their fellow women victimized in the posts. This shows how historical norms embedded in everyday conversations are being transmitted in and through digital spaces. Worse is the way the algorithms are learning and picking these norms and practices that sustain the norms and amplifying them in the digital social spaces.

Some studies point that sexism is a characteristic that increases the interactive nature of social media posts, in fact sexists' posts are more interactive than racial posts (Clarke and Grieve (2017). Misogyny particularly is a major and urgent problem in large social sites like Facebook, and twitter. It includes "*aspects, such as sexual harassment, the stereotypes associated with "stupid" women's behaviour against male, objectification of the female body and a lot of other problems*" (Shushkevich & Cardiff 2021). Our results showed that sexism and misogyny was not only increasing the interactive nature, but that there was an invisible push that was making such posts move faster and get wider engagement. Some sites like twitter have features like high-speed propagation of tweets which not only makes them viral but creates the possibility of these tweets staying in the site for a long time and getting larger viewership through retweeting (Hewitt et al 2016). Facebook on the other hand has features like algorithm reward engagement, which enables the post that receives comments and likes, and other interactions, spread more widely and quickly, being featured more prominently in feeds instead of posts following chronological order of posting.

---

[1] https://www.npr.org/2021/10/05/1043377310/facebook-whistleblower-frances-haugen-congress…

The results of this study show that despite the society being misogynistic, the new algorithmic activity is not only urging this vice on but also amplifying it through post engagement reward systems and a pushed engagement. This has resulted in new values in digital spaces - with online content creators interested more in what content sells as opposed to the content that builds the society. Since sexism and misogyny increases that interactive nature of content, then such content becomes the practice that drives the new norms in the digital spaces. And these norms are directly driven by the AI algorithms entangled in such spaces. Since AI learns from the data that is fed to it every day, then the algorithms pick up these norms and spreads them faster and wider, while learning to perfect this in digital spaces. As a result, new values are not only created, but we see new practices that are propagated by the society entangled in this space as we are going to discuss in the next section.

### 3. Impacts of the new AI ontonorms on Women and Girls

In the African context, posts that seem to drive negative gender norms e.g., misogyny and sexism have been prominent in some social sites, like Facebook and twitter, and have been used by those sites for economic gains. Some studies show that there has been an increase in the content that targets female leaders and influencers[2], mostly presenting them in a negative way. This has not only been used to wash down the gains by female role-modelling, but has also lowered the perceptions of women and girls about their rights in relation to gender based violence. A 2022 demographic and health survey conducted in Kenya shows that 43% of women aged 15-49 believe that a husband is justified to beat his wife. Several reasons are used to support this vice including but not limited to unfaithfulness, coming home late, burning food, going out without reporting to the husband, arguing with the husband, neglecting children and if she refuses to cook[3]. The top reason given for why husbands should beat their wife is unfaithfulness – which our research showed that it could be over-featured and exaggerated in social media, amplified by AI, while at the same time the authenticity of content may not be verified in the era of deep fakes and fictional content creators. While our focus was not necessarily on the posts themselves, and who they target, we argue that this kind of content that involves nudity of female influencers, objectification of the African female body, and amplification of women's 'bad behaviour' and other stereotypes, even in cases where actions or inactions committed involve a male actor - is creating a new oppressive digital spaces, and the data from the content is weaponizing AI to target women and girls unfairly in digital and physical spaces..

---

[2] https://pollicy.org/wp-content/uploads/2023/05/Byte_Bullies_report.pdf
[3] See: https://www.knbs.or.ke/kenya-demographic-and-health-survey-kdhs-2022/

While this is happening, the digital companies in question are amassing massive data, which they use to classify their users as well as create new products that are used to drive their profits (Fourcade and Healy 2017). Since sexist content sells more, this is increasingly attracting new forms of norms and practices among designers and the African digital users - which is further is mediating new forms of digitally gender based violence, with the content emerging from these norms being used by the algorithm to drive profits up for some of the social media sites like twitter and Facebook.

## 5.2 Intentionality in design, training and use

Gender diversity is a very important aspect in technology development because of its ability to draw unique perspectives and knowledge from different genders, and articulate it into designs and other levels of technology development and deployment. It is noteworthy that different gender groups represent certain norms, and the way they do and receive things is different. Our results have not only shown how norms by dominant groups and things are driving new gender norms, but also show how the thriving algorithmic environments are subtly but effectively silencing and subordinating women further, exposing them to social injustices like GBV (whether digitally mediated or actual physical GBV) and other forms of marginalization. Furthermore, the results reveal critical exclusion of women perspectives in AI designs and data in large language models. This implies that AI development and use spaces are characterised by an asymmetrical cognitive environment where women are not deemed as knowledge peers - rather, they are reduced to data givers (lingual, media and behavioral) and just statistics. The concept of gender inclusion AI debates is conveniently being reduced to demographic disaggregation, with most solutions being mere tokenistic additions while muting some aspects of deeper gender perspectives. The norms propagated in these spaces are conveniently being reproduced in the technology itself further marginalizing women and girls in the digital space. This interplay of norms, practices and material technology further entrenches the unequal power relations in the age where AI is being used to classify, commodify and currencify human data for profits (Fourcade and Healy 2017). The key questions are: in the current age of big data are women provided with a conducive environment to engage with AI technology as rational enquirers (Giladi, 2018) or they are being peripheralized, and always forced to seek for epistemic recognition and affirmation from their male peers (Koskinen & Rolin, 2021; Poliseli & Leite, 2021)? Is society busy fixing women using machines? Could that be the reason why the knowledge produced is so weak such that it reduces women to objects of technology (Koskinen & Ludwig, 2021; Ndaka et al., 2024)?. With the current activism against the current rise of GBV, which has been associated with digitally mediated GBV, women have also been subjected with unfair labour of identifying these injustices, and sometimes having to picket to seek for affirmation from dominant groups including the authorities.

We propose intentionality in ensuring inclusion of women and girls in critical spaces that shape how technology is designed, trained, mediated, deployed and governed. Women should not be seen as people who come to manage risk and consequences (Burch et al., 2023; Viseu, 2015) but as groups that carry knowledge contributions that can shape how technology impacts and is impacted by society. Intentionally bringing women into the design space ensures that they influence how and when norms are formed, how activities are done, which practices thrive, as well as ensuring that their strengths are utilized, and the knowledge they carry is acknowledged in a way that recognizes and articulates their needs and values (Felt 2021). Bailey (2022) argues that diversity solves complex human problems, because it brings unique ways of thinking and seeing things - which is critically needed in AI design and use - not just aptitude. He further argues that representation can be a powerful tool that can inspire people conceptualizing AI to strive beyond capabilities and ways of seeing things. That way developed technology will be able to recognize the existence of other worlds (Higgins 2022). This intentionality is not limited to technology design, but also in the way the laws that govern and regulate AI technology are crafted, accountability with individual data, and the algorithmic activities that are taking place in digital spaces. That way, the society will not participate in disgracing the marginalized groups while invisibly and unconsciously enriching powerful tech companies and players.

6.0 Conclusion

In conclusion, this chapter explores the complex effects of AI on various groups, highlighting the new gender norms in digital spaces. The chapter examines how AI algorithms reinforce biased gender norms, with a focus on African women in particular. It also addresses issues like the lack of gender-specific identities, and biased translations in local languages. The paper argues that AI onto-norms shape how AI engages with the content that relates to women in terms of image, behaviour, and other media, which includes how gender identities and perspectives are intentionally or otherwise, included, missed, or misrepresented in building and training AI systems. Drawing from Annemarie Mol's concept of onto-norms, the study uncovers the intricate dynamics between AI social actors, algorithms, and societal norms by studying the nuanced ways in which AI influences the lives of women and girls in Africa. The norms propagated in these spaces, that superiorise male gender and demean female gender in professional and social spaces are conveniently being reproduced in the technology itself further marginalizing women and girls in digital and the society. This interplay of norms, practices and material technology further entrenches the unequal power relations in the age where AI is being used to classify, commodify and currencify human data for profits. Thus this paper underlines the significance of understanding the norms and practices that shape how biases in AI are entrenched. It argues that understanding these norms helps in correcting biases in AI design, training, and application to advance gender equality and mitigate creation of new negative gender norms in and through AI. The paper proposes intentionality in order to ensure the inclusion of women and girls in critical spaces that shape how technology is designed, trained, mediated and deployed, as well as the laws that govern and regulate AI technology, accountability with individual data, and the algorithmic activities that are taking place in digital spaces. That way, the society will not participate, driving the marginalization of already existing groups while invisibly and unconsciously enriching powerful tech companies and players.